\documentclass[preprint,12pt,authoryear=false]{elsarticle}
\usepackage[utf8]{inputenc}
\usepackage[T1]{fontenc}
\usepackage{amsmath,amssymb,graphicx,booktabs,longtable,array,calc,url,hyperref,lineno}
\usepackage{newunicodechar}
\newunicodechar{β}{$\beta$}\newunicodechar{ε}{$\epsilon$}\newunicodechar{θ}{$\theta$}\newunicodechar{γ}{$\gamma$}\newunicodechar{±}{$\pm$}\newunicodechar{×}{$\times$}\newunicodechar{≥}{$\geq$}\newunicodechar{≤}{$\leq$}\newunicodechar{≈}{$\approx$}\newunicodechar{κ}{$\kappa$}\newunicodechar{λ}{$\lambda$}\newunicodechar{μ}{$\mu$}\newunicodechar{σ}{$\sigma$}\newunicodechar{→}{$\to$}\newunicodechar{−}{$-$}\newunicodechar{α}{$\alpha$}\newunicodechar{⁺}{$^{+}$}\newunicodechar{Δ}{$\Delta$}
\providecommand{\tightlist}{\setlength{\itemsep}{0pt}\setlength{\parskip}{0pt}}
\AtBeginEnvironment{longtable}{\footnotesize\setlength{\tabcolsep}{3pt}}
\usepackage{etoolbox}
\biboptions{numbers,sort&compress}
\journal{}

\begin{document}
\begin{frontmatter}
\title{Reducing double-dipping in single-cell topic models with three-way count splitting}
\author[1]{Zheng Qu\corref{cor1}}
\ead{zqu001@mt.feitian.edu}
\cortext[cor1]{Corresponding author}
\affiliation[1]{organization={Fei Tian College Northern Campus},city={Middletown},state={NY},postcode={10940},country={USA}}
\begin{abstract}
Topic models learn gene programs from single-cell RNA sequencing (scRNA-seq) data, and the same counts are then used to test which genes and donors are associated with those programs. Count splitting is the standard remedy for this reuse, but it has not been tested when an encoder reads each cell's counts. On a 104,400-cell lupus cohort with 300 permuted null genes per run, naive tests rejected 36\% of null gene--topic pairs at a nominal 5\% level, and the usual two-way split rejected 38\%, because the encoder sees the tested gene's own held-out count. Encoding topic usage on a separate third of the counts, with beta-binomial thinning, lowered this to 7.7\% (6.7\% for genes outside the model); for two other models, LDVAE and scETM, the rate was 7.0--7.3\% (naive 51--59\%). A label-free interferon program shifts in held-out counts in the lupus cohort and in a pre-specified COVID-19 cohort, and donor-level topic usage predicts disease (AUROC 0.96) beyond cell-type proportions (0.86 and 0.58 in the two cohorts). Calibration and prediction did not depend on the model. We do not claim better clustering, prediction or topic stability. Code, figures and result tables are public (MIT license; DOI 10.5281/zenodo.23307193).
\end{abstract}
\begin{highlights}
\item Naive reuse of counts inflates gene--topic tests (36\% rejected at 5\%)
\item Standard two-way count splitting does not fix this for encoder-based topic models
\item A three-way split with beta-binomial thinning cuts the inflation to about 7\%
\item Label-free interferon program replicates in a pre-specified COVID-19 cohort
\item Donor-level topic usage predicts lupus and COVID-19 status beyond cell types
\end{highlights}
\end{frontmatter}

\section{Introduction}\label{introduction}

Topic models (ProdLDA, scETM, cNMF, scE2TM, Topic-FM) are used to name gene programs in scRNA-seq. Marker tests, enrichment analysis and disease associations on those programs normally reuse the fitting counts (``double dipping''), and published evaluations of such pipelines often rely on clustering scores against labels derived from PCA pipelines and on silhouette scores computed in each method's own embedding. Neither supports claims about the statistical validity of the programs. We make four contributions: (1) a measurement, on real counts with permuted null genes, of how much naive reuse and standard two-way count splitting inflate gene-level tests of a neural topic model (0.36 and 0.38 at nominal 0.05); (2) the identification of the cause (the encoder reads the tested gene's own held-out count), shown for four models, and a three-way split with beta-binomial thinning that removes most of it (0.057--0.077 across models; 0.067 for NTM genes outside the model); (3) a donor-level analysis of program usage in two disease cohorts, with a pre-specified replication; (4) a five-method comparison showing that donor-level calibration and prediction do not depend on the model. We do not claim better clustering, prediction or topic stability.

\subsection{Related work}\label{related-work}

Count splitting divides observed counts into independent parts so that one part can define a structure and another can test it. Poisson count splitting was introduced for inference after latent variable estimation in single-cell data \cite{neufeld2022}; the negative binomial case \cite{neufeld2023a}, data thinning for convolution-closed distributions \cite{neufeld2024} and thinning based on sufficient statistics \cite{dharamshi2025} extend it, and the countsplit R package implements it \cite{neufeld2023b}. These methods give independent halves, but gene-level inference with an encoder-based topic model has a further dependence: the encoder can compute each held-out cell's topic usage from the tested gene's own held-out count (Section 3.2). We therefore split counts three ways, for fitting, for encoding topic usage and for testing.

This dependence is a form of double dipping: data used to define a structure can also influence tests of that structure. Selective inference for hierarchical clustering \cite{gao2022}, valid post-clustering differential analysis \cite{zhang2019} and ClusterDE \cite{song2023,lee2026} address related concerns. Our focus is narrower: rather than correcting inference after choosing clusters, we isolate the tested gene from the encoder's estimate of cell-level topic usage.

Autoencoding variational inference has been developed for topic models \cite{srivastava2017}; related single-cell approaches include interpretable cellular and gene-signature embeddings \cite{zhao2021}, gene-expression programs \cite{kotliar2019}, and interpretable factor models via variational autoencoders \cite{svensson2020}. Other latent-variable approaches include scVI \cite{lopez2018} and hierarchical Poisson factorization \cite{levitin2019}. These models are used for gene-level interpretation, but good representation or prediction does not make gene-level tests valid.

In multi-sample cohorts, donor-level inference treats donors, not cells, as the experimental units: counts can be aggregated by donor and cell population for pseudobulk testing. Pseudobulk methods for multi-sample data \cite{crowell2020} and the evidence of false discoveries in cell-level differential expression \cite{squair2021} motivate this. Donor-level testing and count splitting address different sources of uncertainty; the latter targets leakage from an encoder into a gene-level test, including settings where topic usage is estimated from the same cells being tested.

\section{Methods}\label{methods}

\textbf{Model.} For cell \emph{i}: z\_i \textasciitilde{} N(μ(x\_i), σ²(x\_i)), θ\_i = softmax(z\_i), p\_i = softmax(θ\_i B + b + γ\_\{c(i)\}), x\_i \textbar{} n\_i \textasciitilde{} Multinomial(n\_i, p\_i). B (K×G) are topic--gene logits; b is a free gene background (the role TF-IDF was meant to play); γ\_\{c\} are optional intercepts for a categorical covariate c(i) (donor in Kang; cell type in lupus, so identity programs are absorbed and topics describe within-type programs; donor intercepts are not used in the lupus analysis because they absorb donor-level disease effects). Encoder input is log(1+1000·x/n). Training uses Adam and KL annealing.

\textbf{Count splitting.} Each UMI count is split x = x\^{}(tr) + x\^{}(te) with x\^{}(tr) \textbar{} x \textasciitilde{} Binomial(x, ε) (independent halves under Poisson) or BetaBinomial(x, εr, (1−ε)r) (independent halves under NB with size r; data thinning). ε = 0.5 throughout. Every data-dependent step (gene filtering, HVG selection, model fitting, labeling topics by enrichment) uses x\^{}(tr) only. Inference applies the frozen encoder to x\^{}(te) and tests on x\^{}(te).

\textbf{Three-way split.} Counts are split into thirds by two (beta-)binomial thinning steps: the first third fits the model (and every data-dependent choice), the second is passed through the frozen encoder to obtain program usage, and gene-level tests relate that usage to the third's counts. Under Poisson (or NB with the right size) the three parts are independent, so a cell's usage no longer depends on the counts being tested. The two-way split used elsewhere in this paper evaluates usage and expression on the same held-out counts.

\textbf{Data partitions.} Model fitting uses only the training part of each split (one half in the two-way split, the first third in the three-way split); every gene-level test uses disjoint held-out counts of the same cells. Donor-level prediction is evaluated by 5×5 cross-validation over donors, and a fully nested version refits every data-dependent step inside each training fold (Section 3.4).

\textbf{Tests.} (a) Gene--program association: partial correlation of log-normalized held-out expression with program usage adjusting for depth (and donor and cell type in real data). (b) Donor level: per-donor mean program usage within a cell type; paired t-test (Kang) or covariate-adjusted OLS with HC3 errors (lupus; age, sex, ethnicity groups with ≥10 donors). (c) Program labeling: hypergeometric enrichment of the 50 most specific genes (β\_k / mean β) in Hallmark IFN-α/γ sets (universe: modeled HVGs), from training-half β only.

\textbf{Simulation (E1).} 3,000 cells, 4 topics, 300 genes (100 informative, 200 null); Poisson or NB counts (size 2 and 0.5), mean depth 1 and 0.3; null genes independent of θ given size factor; library size independent of θ by construction. 10 seeds per scenario. ``Naive'' fits and tests on all counts. Rates are over null gene × topic pairs.

\textbf{Null genes on real data (E3b).} In the lupus subsample, 300 real genes (expressed in ≥5\% of cells; random per seed) are made null by permuting their counts across cells within each donor × cell-type stratum, then appended to the matrix and kept in the model. Tests use partial correlation with stratum means and log library size projected out, 10 seeds, NTM-ct K = 30 (\texttt{code/e3\_nullgenes.py}).

\textbf{Baselines (E3).} Under the identical protocol (same split, K = 30, seeds 0--9, same donor-level tests and CV) we fit NMF, scVI's LDVAE, scETM (100 epochs, defaults) and cNMF (20 iterations). Baselines are untuned and run on CPU; the NTM ran on a GPU. For every method the program is chosen by best-of-K label-free ISG enrichment on the training half, which is optimistic and applied equally.

\textbf{Real data.} \emph{E2:} GEO GSE96583 batch 2 \cite{kang2018}: 24,132 singlets after removing 313 barcodes shared by both conditions; 8 donors, control and IFN-β; K = 20, donor intercepts, 10 seeds. Baselines under the same split: NMF (log-normalized) and scikit-learn LDA (counts). \emph{E3:} Perez et al.~\cite{perez2022} SLE PBMC (the full study has 1,263,676 cells from 264 samples; we use CELLxGENE Census; 261 donors, 162 SLE, 99 controls), donor-stratified subsample of 400 cells per donor (104,400 cells); K = 30; 10 seeds; 5×5 cross-validated L2 logistic regression for donor-level disease prediction with features from the held-out half only. Feature sets: covariates (age, sex, ethnicity), cell-type proportions, per-donor×cell-type mean topic usage, and per-donor×cell-type mean of 30 PCs (pseudobulk baseline). Topics are learned without labels; the main analysis is transductive (topic learning sees all donors), and a fully nested version refits everything inside each training fold (\texttt{code/e3\_nested.py}).

\section{Results}\label{results}

\subsection{Naive reuse of counts inflates gene--program tests; two-way count splitting restores calibration in simulation (E1; Fig. 1)}\label{naive-reuse-of-counts-inflates-geneprogram-tests-two-way-count-splitting-restores-calibration-in-simulation-e1-fig.-1}

Naive false-positive rates were 0.38--0.49 across all five scenarios (nominal 0.05). Binomial count splitting gave 0.049--0.062 under Poisson and mild NB but 0.082 (size 2, depth 1) and 0.154 (size 0.5) under strong overdispersion. NB-aware splitting gave 0.054--0.061 in every scenario. Split tests had much lower power than naive tests (0.07--0.29 for an informative gene at the Bonferroni-over-topics level); this power loss reflects the cost of validity and was largest at low depth (power versus ε and a quasi-Poisson score test: Supplement S2).

\begin{figure}[t]\centering\includegraphics[width=\linewidth]{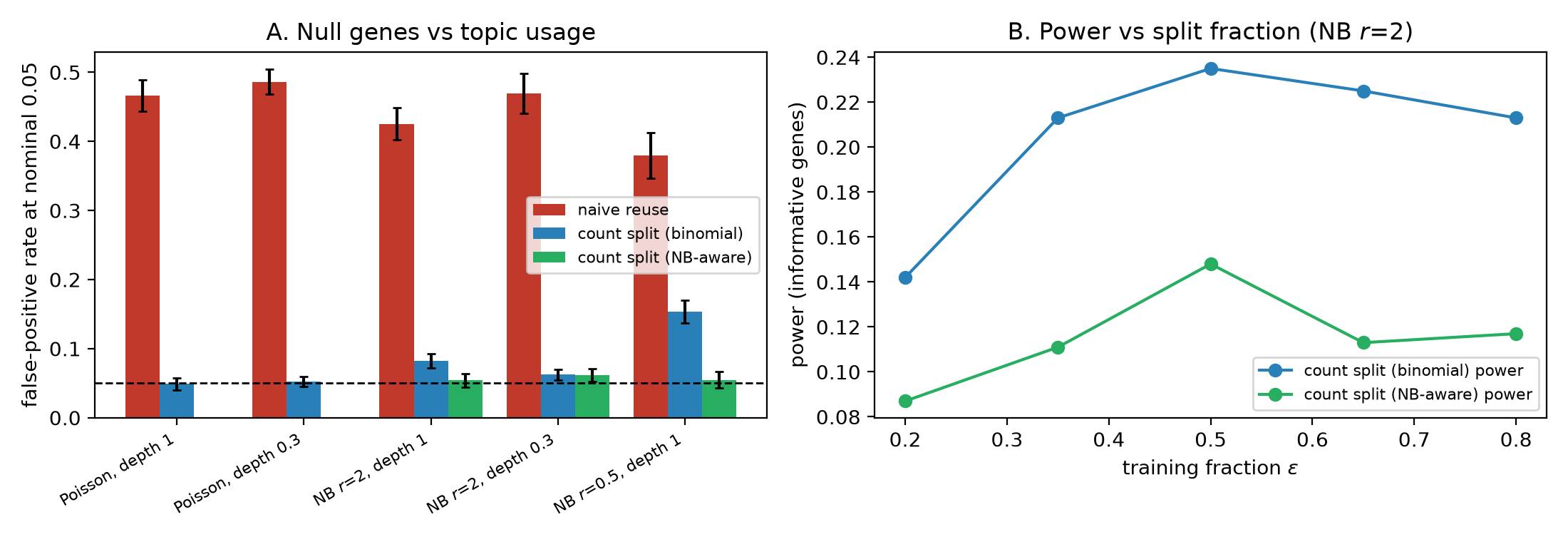}
\caption{Simulation (E1; 3,000 cells, 4 topics, 300 genes, 10 seeds). (A) False-positive rate at nominal 0.05 for null genes tested against topic usage: naive reuse, binomial count splitting and NB-aware (beta-binomial) splitting across Poisson and NB scenarios. (B) Power for informative genes versus the training fraction ε (NB size 2).}\label{fig:1}\end{figure}

\subsection{Real-count null genes: the standard split does not remove gene-level inflation, the three-way split removes most of it (E3b)}\label{real-count-null-genes-the-standard-split-does-not-remove-gene-level-inflation-the-three-way-split-removes-most-of-it-e3b}

Table 1 gives the rejection rate among null gene × topic pairs at nominal 0.05. For NTM-ct, naive reuse rejected 0.360 ± 0.021 and the two-way split 0.379 ± 0.032 (0.246 and 0.265 at nominal 0.01). The three-way binomial split gave 0.104 ± 0.007 (0.050 at 0.01) and the three-way beta-binomial split 0.077 ± 0.005 (0.027). For genes outside the model, the two-way split gave 0.085 ± 0.007 and the three-way beta-binomial split 0.067 ± 0.006 (0.018 at 0.01). The test itself, with usage shuffled among cells within stratum, rejected 0.047--0.061. The two-way split fails because the encoder derives each held-out cell's usage from the tested gene's own held-out count, not because the halves are dependent. Even the best NTM protocol is above nominal, so we claim a large reduction, not exact calibration.

Other models showed the same pattern (Table 1; Fig. 2). With NMF on log-normalized counts the three-way split came close to nominal (0.057 ± 0.003; 0.016 at 0.01); for the three neural encoders it left a residual of 0.07--0.08. Marker genes of the lupus ISG topic in classical monocytes show the same ordering (non-ISG genes called: naive 0.159, two-way 0.085, three-way 0.048), while top-100 ISG precision is kept (0.164). Limits: permutation within strata tests dependence through the model, not biological confounding; NB sizes are estimated on all counts; the check covers four models and one cohort, and cNMF was not run (its usage step resembles NMF's).

\textbf{Table 1.} Rejection rate among permuted null gene × topic pairs at nominal 0.05 (lupus, 104,400 cells; mean ± SD over seeds). NMF used K = 30 on log-normalized counts. LDVAE and scETM ran on CPU with 5 seeds.

\begin{longtable}[]{@{}lllll@{}}
\toprule\noalign{}
Model & Seeds & Naive & Two-way split & Three-way beta-binomial \\
\midrule\noalign{}
\endhead
\bottomrule\noalign{}
\endlastfoot
NTM-ct & 10 & 0.360 ± 0.021 & 0.379 ± 0.032 & 0.077 ± 0.005 \\
NMF & 10 & 0.251 ± 0.021 & 0.173 ± 0.014 & 0.057 ± 0.003 \\
LDVAE & 5 & 0.593 ± 0.005 & 0.556 ± 0.031 & 0.073 ± 0.003 \\
scETM & 5 & 0.505 ± 0.026 & 0.571 ± 0.031 & 0.070 ± 0.004 \\
\end{longtable}

\begin{figure}[t]\centering\includegraphics[width=\linewidth]{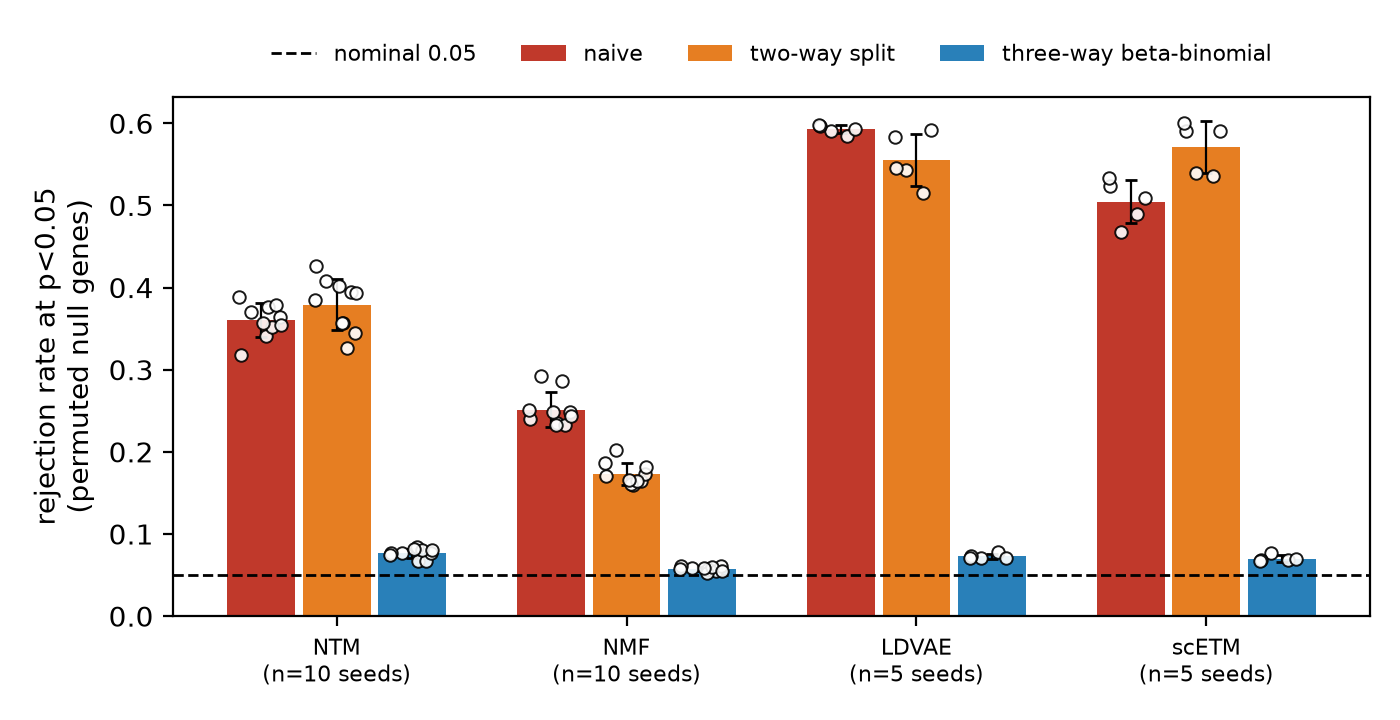}
\caption{Real-count null genes (E3b; lupus, 104,400 cells, 300 permuted genes per run). Rejection rate at nominal 0.05 (dashed line) by protocol (naive, two-way split, three-way beta-binomial; the three-way binomial is in Section 3.2) and model (NTM-ct, NMF, LDVAE, scETM); NTM-ct and NMF use 10 seeds, LDVAE and scETM 5.}\label{fig:2}\end{figure}

\subsection{IFN-β PBMCs: a label-free interferon program replicates in held-out counts (E2; Fig. 3)}\label{ifn-ux3b2-pbmcs-a-label-free-interferon-program-replicates-in-held-out-counts-e2-fig.-3}

Over 10 seeds the topic most enriched for Hallmark IFN genes (median enrichment p = 2.2e-13) was identified from training counts alone; its held-out usage increased with stimulation in 10/10 seeds (paired over 8 donors; mean usage change +0.159 ± 0.031; CD14⁺ monocytes +0.515 ± 0.130; FCGR3A⁺ monocytes +0.379 ± 0.076; lymphocytes ≤0.05). NMF and LDA fit under the same protocol recover a comparable program (median enrichment p 2.5e-16 and 1.0e-19), so the IFN signal is a positive control rather than a discriminating benchmark. Topic stability across seeds (best-match Jaccard of top-50 specific genes over 683 shared genes) was equal across methods (NTM 0.485, NMF 0.458, LDA 0.490). Among marker genes of the IFN program, 19.4\% ± 1.7\% of non-ISG genes were called at FDR 5\% with held-out counts versus 34.9\% ± 4.6\% with naive reuse (median non-ISG p 0.252 vs 0.100); recovered ISGs were equal (24.4 vs 23.9). Since many non-ISG genes do respond to IFN-β, this contrast is supportive, not calibrating; calibration is established by E1. Clustering against author cell types (KMeans, 8 types): NTM θ ARI 0.467 ± 0.047 and NMI 0.621 ± 0.019; PCA-30 ARI 0.481 ± 0.013 and NMI 0.628 ± 0.010, i.e.~comparable.

\begin{figure}[t]\centering\includegraphics[width=\linewidth]{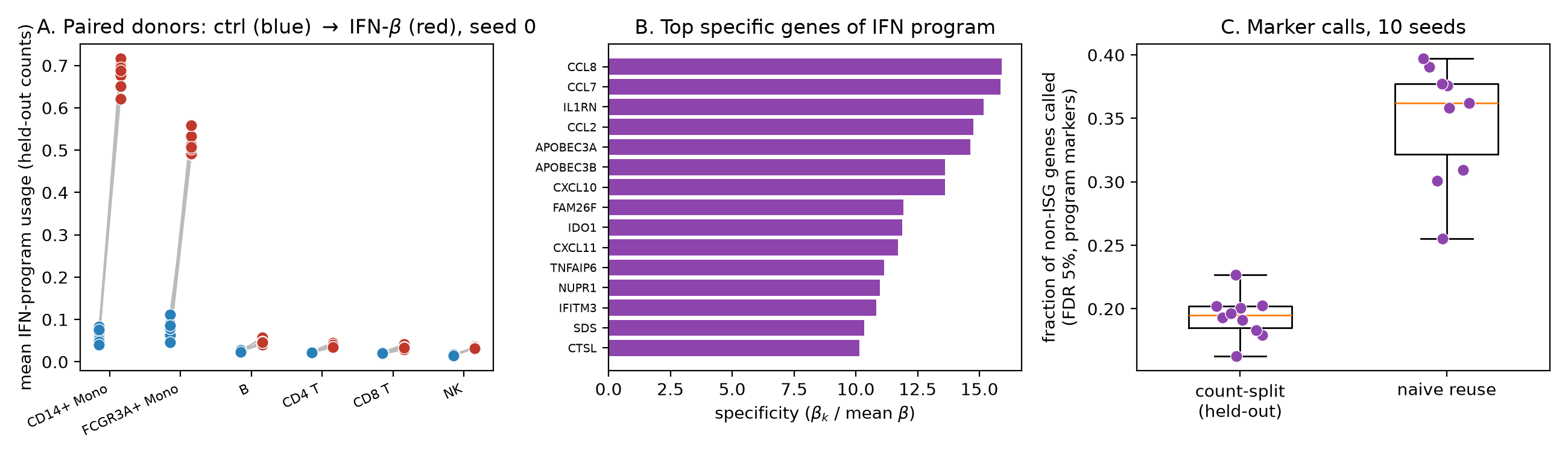}
\caption{IFN-β PBMCs (E2; Kang et al., 8 donors). (A) Mean held-out usage of the label-free IFN program per donor, control versus IFN-β (seed 0). (B) Top specific genes of that program. (C) Fraction of non-ISG genes called as program markers at FDR 5\%, count-split versus naive reuse (10 seeds).}\label{fig:3}\end{figure}

\subsection{Lupus cohort: a pan-cell-type interferon program and donor-level prediction (E3; Fig. 4)}\label{lupus-cohort-a-pan-cell-type-interferon-program-and-donor-level-prediction-e3-fig.-4}

\emph{Program discovery.} With cell-type intercepts, the topic most enriched for IFN genes (median p = 1.5e-11) is dominated by IFI27, ISG15, HERC5, IFIT2, CXCL10 in the three seeds whose top genes we inspected. Its held-out usage was higher in SLE donors after adjusting for age, sex and ethnicity in CD4 T (0.70 SD, 10/10 seeds with p\textless0.05), classical monocytes (0.80 SD, 10/10), CD8 T (0.51 SD, 9/10), B (0.60 SD, 9/10), non-classical monocytes (0.89 SD, 10/10) and cDCs (0.90 SD, 8/10), and weakly in NK cells (0.24 SD, 5/10). The program is therefore shared across lineages with cell-type-specific magnitude, which a cluster-based analysis would split into separate states.

\begin{figure}[t]\centering\includegraphics[width=\linewidth]{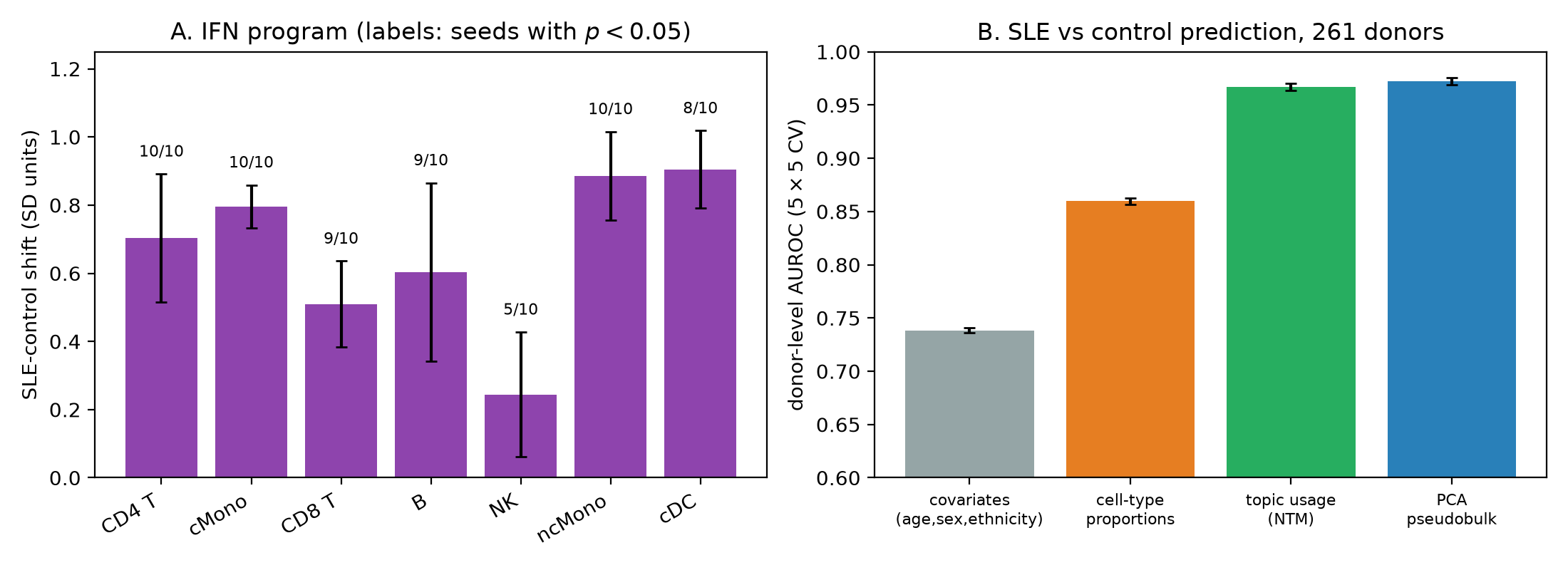}
\caption{Lupus cohort (E3; 261 donors). (A) SLE-minus-control shift in held-out usage of the IFN program by cell type, in SD units (labels: seeds with p\textless0.05 of 10). (B) Donor-level AUROC (5×5 CV) for SLE versus control from covariates, cell-type proportions, topic usage and PCA pseudobulk.}\label{fig:4}\end{figure}

\emph{Calibration on real data.} Permuting disease labels among donors (10 permutations × 210 topic×cell-type tests × 10 seeds) gave rejection rates of 0.043 at 0.05 and 0.0087 at 0.01 (slightly conservative). This holds for any label-free topic analysis tested at the donor level, since topics do not depend on the labels; it shows that the donor-level inference is sound, not that splitting is needed for it. The role of splitting is at the gene level (Section 3.2).

\emph{Prediction.} Donor-level AUROC (5×5 CV, mean ± SD over 10 seeds): covariates 0.738 ± 0.002; cell-type proportions 0.860 ± 0.003; topic usage 0.967 ± 0.004; PCA pseudobulk 0.972 ± 0.004. Topic usage beat cell-type proportions by 0.107 AUROC in 10/10 seeds, and adding topic usage to proportions + covariates raised AUROC from 0.877 to 0.975. \emph{Fully nested analysis.} Refitting gene filtering, HVGs and the topic model inside each outer training fold (test donors' cells never seen; 5 donor folds, 10 seeds) gave AUROC 0.964 ± 0.004 for topics, 0.971 ± 0.006 for PCA pseudobulk, 0.860 ± 0.005 for proportions and 0.740 ± 0.007 for covariates; topics beat proportions by 0.104 in 10/10 seeds, so the transductive result (0.967) was not inflated beyond about 0.003. Topic usage did \textbf{not} beat PCA pseudobulk (−0.005; PCA ahead in 9/10 seeds); the claim is that interpretable programs carry as much donor-level disease information as an uninterpretable embedding, not more.

\emph{Naive versus split markers.} For the ISG topic in classical monocytes (25,605 cells), naive reuse called 208 ± 53 marker genes at FDR 5\% versus 116 ± 70 with held-out counts, and called 15.9\% ± 4.7\% of non-ISG genes versus 8.5\% ± 5.1\% (naive higher in 10/10 seeds); in CD4 T cells 16.3\% versus 11.5\% (9/10). Naive reuse also recovered more ISGs (25.8 vs 13.6), so it gains power and inflation together; as in E2 the contrast is supportive, not calibrating.

\emph{Supplementary analyses.} Sensitivity to K (15, 50), the beta-binomial split on real lupus counts, and the ablation without cell-type intercepts are in Appendix S1; in brief, calibration and prediction were unchanged across K = 15, 30, 50 (null rejection 0.047--0.052, AUROC 0.963--0.972), and removing cell-type intercepts weakened enrichment (median p 1.9e-4) and cross-lineage detection but not prediction (AUROC 0.962).

\emph{External replication (COVID-19 PBMC; hypotheses and analysis fixed in a written plan before the data were analyzed).} In the Stephenson et al.~\cite{stephenson2021} cohort (115 donors: 86 COVID-19, 29 healthy; 46,000 cells; same protocol, no tuning) all three pre-specified hypotheses held: permutation-null rejection 0.053 at 0.05; the most IFN-enriched topic (median enrichment p 7.6e-13) shifted in COVID-19 in held-out counts in classical monocytes (0.73 SD, 9/10 seeds), CD4 T, CD8 T and NK cells (10/10 seeds each); and topic usage predicted COVID-19 status (AUROC 0.949 ± 0.011) far better than cell-type proportions (0.581 ± 0.011; 10/10 seeds). PCA pseudobulk was again slightly ahead (0.961). Age is strongly confounded with disease in this cohort (covariates alone 0.729).

\emph{Baselines (Table 2).} Calibration was not specific to the neural model: all four baselines were also calibrated under donor-label permutation, and prediction differences were small (LDVAE 0.979 \textgreater{} scETM 0.975 \textgreater{} PCA 0.972 \textgreater{} NTM-ct = cNMF 0.967 \textgreater{} NMF 0.962; significance of the gaps was not tested). NTM-ct had the strongest ISG enrichment, with cNMF second; but NTM-ct uses cell-type labels as intercepts and the baselines do not, while the label-free NTM (median p 1.9e-4) was weaker than cNMF, NMF and LDVAE, so this is not a like-for-like ranking. We therefore claim calibration as a property of the protocol and do not claim better prediction.

\textbf{Table 2.} Lupus cohort, five methods under the identical protocol (K = 30, seeds 0--9). Rows: donor-label permutation rejection at p\textless.05 and p\textless.01; donor-level AUROC from per-donor × cell-type mean topic usage; median p of best-of-K Hallmark IFN enrichment on the training half. NTM-ct uses cell-type intercepts, the baselines do not; baselines untuned.

\begin{longtable}[]{@{}
  >{\raggedright\arraybackslash}p{(\columnwidth - 10\tabcolsep) * \real{0.1667}}
  >{\raggedright\arraybackslash}p{(\columnwidth - 10\tabcolsep) * \real{0.1667}}
  >{\raggedright\arraybackslash}p{(\columnwidth - 10\tabcolsep) * \real{0.1667}}
  >{\raggedright\arraybackslash}p{(\columnwidth - 10\tabcolsep) * \real{0.1667}}
  >{\raggedright\arraybackslash}p{(\columnwidth - 10\tabcolsep) * \real{0.1667}}
  >{\raggedright\arraybackslash}p{(\columnwidth - 10\tabcolsep) * \real{0.1667}}@{}}
\toprule\noalign{}
\begin{minipage}[b]{\linewidth}\raggedright
Metric (10 seeds, mean ± SD)
\end{minipage} & \begin{minipage}[b]{\linewidth}\raggedright
cNMF
\end{minipage} & \begin{minipage}[b]{\linewidth}\raggedright
scETM
\end{minipage} & \begin{minipage}[b]{\linewidth}\raggedright
LDVAE
\end{minipage} & \begin{minipage}[b]{\linewidth}\raggedright
NMF
\end{minipage} & \begin{minipage}[b]{\linewidth}\raggedright
NTM-ct
\end{minipage} \\
\midrule\noalign{}
\endhead
\bottomrule\noalign{}
\endlastfoot
Permutation-null rejection, p\textless.05 & 0.0470 ± 0.0056 & 0.0480 ± 0.0061 & 0.0471 ± 0.0064 & 0.0442 ± 0.0047 & 0.043 \\
Permutation-null rejection, p\textless.01 & 0.0096 ± 0.0029 & 0.0093 ± 0.0012 & 0.0088 ± 0.0028 & 0.0076 ± 0.0015 & 0.0087 \\
AUROC, topic pseudobulk (5×5 CV) & 0.967 ± 0.004 & 0.975 ± 0.003 & 0.979 ± 0.003 & 0.962 ± 0.005 & 0.967 ± 0.004 \\
ISG enrichment, median p (best-of-K) & 7.9e-9 & 0.034 & 1.1e-6 & 1.3e-7 & 1.5e-11 \\
\end{longtable}

\section{Discussion and limitations}\label{discussion-and-limitations}

\begin{itemize}
\tightlist
\item
  Cell-level gene tests treat cells as exchangeable and are liberal in real data; donor-level tests are the inferential unit. The E1 guarantee covers gene-level tests under Poisson/NB models; real counts depart from both, and NB size must be estimated.
\item
  Count splitting halves information per fold; for rare programs, power can be low (Fig. 1B).
\item
  Topic selection by \textbar t\textbar{} across 8 donors favors consistent-but-tiny shifts; we label programs by label-free enrichment instead.
\item
  Kang is a positive control only; NMF and LDA find the same program. Clustering accuracy is not the claim.
\item
  The main prediction analysis is transductive; the nested analysis (Section 3.4) gives nearly identical results. SLE vs control is strongly confounded with sex in this cohort; covariates are adjusted but residual confounding is possible. A second cohort (COVID-19) replicates the main findings, but both are PBMC datasets of adults analyzed with the same pipeline.
\item
  Cell-type intercepts require labels; they are used only to absorb identity and are not used as outcomes.
\item
  Baselines are untuned (scETM 100 epochs, cNMF 20 iterations), selected best-of-K, run on CPU while the NTM ran on a different device, and effect sizes in SD are not comparable across models.
\item
  The E1 simulation (3,000 cells) shows the two-way split calibrated, but E3b shows it is not at 104,400 cells on real counts; E1 has too little power to see the encoder's dependence on the tested gene. The three-way split is still 1.5× nominal in E3b and spends two thirds of the counts on fitting and encoding.
\item
  The seed counts (10/10 etc.) measure variation from the split and initialization on a fixed donor set, not sampling variation of donors; no donor-level confidence intervals are reported for AUROC gaps.
\item
  Cell-type labels come from the dataset authors' annotation, made on the same data, and serve as strata and intercepts, not as outcomes.
\item
  Not done: the E3b check for cNMF, donor-bootstrap CIs, tuned baselines.
\end{itemize}

\section*{Data and code availability}
Kang et al.\ data: GEO GSE96583 (batch 2) \cite{kang2018}. Perez et al.\ lupus data \cite{perez2022}: CELLxGENE Census dataset 218acb0f-9f2f-4f76-b90b-15a4b7c7f629 (stable release 2025-11-08). Stephenson et al.\ COVID-19 data \cite{stephenson2021}: CELLxGENE Census dataset c7775e88-49bf-4ba2-a03b-93f00447c958. Analysis code, figures and result tables are released under the MIT license at \url{https://github.com/quz2004/topic-model-count-splitting} (release v0.11, archived at Zenodo: \url{https://doi.org/10.5281/zenodo.23307193}).

\section*{Reproducibility}
Seeds 0--9 throughout; the same seed gives the same count split across methods. Software versions, run manifests and seed conventions are in Appendix S3. 

\section*{CRediT authorship contribution statement}
\textbf{Zheng Qu:} Conceptualization, Methodology, Software, Formal analysis, Investigation, Data curation, Visualization, Writing -- original draft, Writing -- review \& editing.

\section*{Declaration of competing interest}
The author declares no known competing financial interests or personal relationships that could have appeared to influence the work reported in this paper.

\section*{Funding}
This research did not receive any specific grant from funding agencies in the public, commercial, or not-for-profit sectors.

\section*{Declaration of generative AI and AI-assisted technologies in the writing process}
During the preparation of this work the author used Claude (Anthropic, via Claude Code) and GPT-based models to help write analysis code, run analyses and edit the manuscript text. After using these tools, the author reviewed and edited the content as needed and takes full responsibility for the content of the publication.

\bibliographystyle{elsarticle-num}
\bibliography{refs}

\appendix
\renewcommand{\thesection}{S\arabic{section}}
\setcounter{section}{0}
\section*{Supplementary material}
\section{S1 Additional lupus analyses}\label{s1-additional-lupus-analyses}

\emph{Sensitivity to K.} At K = 15 and 50 (10 seeds each) the permutation-null rejection rate stayed at 0.052 and 0.047 (nominal 0.05), topic-usage AUROC was 0.972 and 0.963 (PCA 0.972), and topics beat cell-type proportions by 0.113 and 0.103 in 10/10 seeds. Label-free ISG enrichment was weaker at K = 15 (median p 1.3e-5) and similar at K = 50 (2.1e-12); at K = 50 the IFN signal is split across topics and the best single topic is detected less consistently in B (5/10 seeds) and NK cells (2/10).

\emph{Beta-binomial split on real counts.} Per-gene NB sizes estimated by method of moments (median 2.2 across expressed genes) were used for beta-binomial splitting: null rejection 0.049 at 0.05, topic AUROC 0.968, ISG enrichment median p 1.7e-13; the IFN shift stayed significant in 10/10 seeds in classical and non-classical monocytes but fell to 8/10 (CD4 T, CD8 T), 6/10 (B) and 5/10 (cDC). Caveat: sizes were estimated on all counts.

\emph{Ablation (no cell-type intercepts, 10 seeds).} The most IFN-enriched topic mixes identity and IFN genes (median enrichment p = 1.9e-4 vs 1.5e-11) and its SLE shift is detected consistently only in monocytes and CD4/CD8 T cells (cMono 0.33 SD, 10/10 seeds; ncMono 0.52 SD, 10/10; CD8 T 0.35 SD, 7/10; CD4 T 0.25 SD, 9/10) but not B cells (0.07 SD, 2/10), NK (0.15 SD, 3/10) or cDCs (0.26 SD, 1/10). Prediction is nearly unchanged (topic usage AUROC 0.962 ± 0.006) and label-permutation calibration is similar (0.053 at 0.05). Cell-type intercepts therefore matter for interpretability and cross-lineage detection, not for predictive accuracy.

\section{S2 Additional simulation results (E1)}\label{s2-additional-simulation-results-e1}

Split-test power was flat in ε between 0.2 and 0.8. A quasi-Poisson score test had slightly higher power than correlation at depth 1 (0.19 vs 0.16 in a 3-seed check) with calibrated error.

\section{S3 Reproducibility details}\label{s3-reproducibility-details}

Seeds 0--9. Laptop runs: Python 3.14, PyTorch 2.14 (MPS); baselines and later runs on DGX Spark: Python 3.12, PyTorch 2.13 (CUDA), scvi-tools 1.5.1, cnmf 1.7.1, scetm 0.5.0. Each run is recorded in a manifest (exact command, seed, git commit, host, device, library versions, output files); runs made before the manifest existed were reconstructed from result file names and are marked as such. Count split: \texttt{numpy.random.default\_rng(s)}; model initialization, PCA and cross-validation splitters use the same seed s (nested CV fold f uses 10s+f). GPU nondeterminism was not controlled, and CUDA reruns do not reproduce MPS runs bit for bit. Replication cohort: 400 cells per donor (\texttt{code/fetch\_covid.py}).

\end{document}